\documentstyle[emulateapj, psfig]{article}
\begin{document}
\title{The Nature of the UV/X-Ray Absorber in PG~2302+029}
\author{Bassem M. Sabra \& Fred Hamann}
\affil{Department of Astronomy, University of Florida, Gainesville, 
FL 32611}
\author{Buell T. Jannuzi}
\affil{National Optical Astronomy Observatory, 950 North Cherry Ave., Tuscon, 
AZ 85719}
\author{Ian M. George}
\affil{Joint Center for Astrophysics, Department of Physics, 
University of Maryland, Baltimore County, 1000 Hilltop Circle, Baltimore, 
MD 21250}
\affil{Laboratory for High Energy Astrophysics, Code 662, 
NASA/Goddard Space Flight Center, Greenbelt, MD 20771}

\author{Joseph C. Shields}
\affil{Department of Physics \& Astronomy, Ohio University, Athens, OH 45701}
\submitted{Accepted for Publication in the Astrophysical Journal}

\begin{abstract}
We present {\it Chandra} X-ray observations of the radio-quiet QSO 
PG~2302+029. This quasar has a rare system of ultra-high velocity 
($-56,000$ km s$^{-1}$) UV absorption lines that form in an outflow from the 
active nucleus (Jannuzi et al. 2003). The {\it Chandra} data indicate that 
soft X-ray absorption is also present. We perform a joint UV and X-ray 
analysis, using photoionization calculations, to detemine the nature of 
the absorbing gas. The UV and X-ray datasets were not obtained simultaneously. 
Nonetheless, 
our analysis suggests that the X-ray absorption occurs at high velocities 
in the same general region as the UV absorber. There are not enough 
constraints to rule out multi-zone models. In fact, the distinct broad and 
narrow UV line profiles clearly indicate that multiple zones are present. 
Our preferred estimates of the ionization and total column density in 
the X-ray absorber ($\log U=1.6$, $N_{\rm H}=10^{22.4}$ cm$^{-2}$) 
over predict the \ion{O}{6} $\lambda \lambda 1032, 1038$ absorption 
{\it unless} 
the X-ray absorber is also outflowing at $\sim 56,000$ km s$^{-1}$, but they  
over predict the \ion{Ne}{8} $\lambda \lambda 770, 780$ absorption at 
{\it all} 
velocities. If we assume that the X-ray absorbing gas is outflowing 
at the same velocity of the UV-absorbing wind and that the wind 
is radiatively accelerated, then the outflow must be launched at a radius of  
$\le 10^{15}$ cm from the central continuum source. The smallness 
of this radius casts doubts on the assumption of radiative acceleration. 
\end{abstract}

\keywords{galaxies: active---quasars: absorption lines---quasars: 
individual (PG~2302+029)---X-rays: galaxies}

\section{Introduction}

X-ray absorption in quasars provides a powerful tool to study QSO 
environments. The signatures of this absorption are typically the suppression 
of soft X-rays and /or the presence of absorption edges from \ion{O}{7} 
and \ion{O}{8} near 0.8 keV (Reynolds 1997; George et al. 1998). The 
strength of these features depends mainly on the degree of ionization 
and total column density. The relationship of the X-ray absorber to other 
components of the quasar environs, especially the UV absorbing gas, is not 
well understood. A natural question is whether these features originate in the 
same gas. 

The best-known intrinsic UV absorption lines in QSOs are the broad 
absorption lines (BALs), but recent work has shown that some of the observed 
narrow absorption lines (NALs) and ``mini-BALs'' 
are also intrinsic to QSO environments 
(Barlow, Hamann \& Sargent 1997; Hamann et al. 1997a and 1997b). These 
features can have 
velocity shifts comparable to the BALs (up to $-$51,000~km~s$^{-1}$ 
in one confirmed case; Hamann et al. 1997b) but the narrow line widths 
(from $<$100~km~s$^{-1}$ to a few thousand km~s$^{-1}$) require outflows 
with much smaller line-of-sight velocity dispersions. BALs, mini-BALs, 
and intrinsic 
NALs are each rare in QSO spectra, but the outflows that cause them might 
be common if, as expected, the gas covers a small fraction of the sky as 
seen from the central source.  Among Seyfert 1 galaxies, 50\% 
show 
intrinsic UV absorption (Crenshaw et al. 1999), and there is also a one-to-one 
correlation between the detection of X-ray absorption in Seyfert 1 galaxies 
and the appearance of intrinsic UV absorption lines (Crenshaw et al. 1999).  

In QSOs, X-ray absorption is much rarer  
(Laor et al. 1997). Brandt, Laor, \& Wills (2000)
showed that QSOs with BALs are all significantly weaker soft X-ray sources 
than comparable-luminosity QSOs without BALs. Positive X-ray detections of 
BALQSOs 
(e.g., Mathur, Elvis, \& Singh 1995;  Green et al. 2001; Sabra \& Hamann 2001; 
Gallagher et al. 2002) reveal large absorbing columns of 
$N_{\rm H} \ga 10^{23}$~cm$^{-2}$. QSOs with weaker intrinsic UV absorption 
lines, e.g., the NALs and mini-BALs, appear to have systematically weaker 
X-ray absorption (Brandt et al. 2000). However, more work is needed to 
characterize the X-ray absorption in quasars with different types/strengths
of UV absorption. Studies of X-ray absorbers in 
quasars, especially in relation to the UV 
absorption lines, will have profound impact on our knowledge of quasar 
wind properties, such as the acceleration mechanism, outflow geometry, 
wind launch radius, and mass loss rate (e.g., Mathur et al. 1995; 
Murray, Chiang, \& Grossman 1995; Hamann 1998; Sabra \& Hamann 2001). 
For example, it is difficult to radiatively accelerate an outflow with a 
large column density, unless it is launched from the inner most regions of 
the accretion disk.

In this paper we will discuss \textit{Chandra} X-ray observations of the QSO 
PG~2302+029. This object shows a system of intrinsic UV absorption lines 
that have a velocity shift of $-56,000$~km~s$^{-1}$ (Jannuzi et al. 1996) 
with respect to the systemic velocity of the 
QSO ($z_{ems}=1.044$). The system consists of NALs 
($FWHM \approx 330$ km~s$^{-1}$) at $z=0.7016$ and a ``mini-BAL'' 
system ($FWHM \approx 3300$ km~s$^{-1}$) at $z=0.695$ (Jannuzi et al. 
1996; 1998). It is also characterized by being an X-ray faint source  with 
an {\it  HEAO-1} A-2 flux upper limit at 2~keV of 
$4\times 10^{-12}$~erg~s$^{-1}$~cm$^{-2}$ (Della Ceca et al. 1990). 
Our aim is to determine the properties of the X-ray spectrum, search for 
signs of absorption, and define the relationship between the UV and X-ray 
absorbing gas. We also discuss the location of the absorber, and the 
potential implications of high-velocity X-ray absorption for the wind 
dynamics.  
If the X-ray and UV absorbers are the same, then the $-56,000$~km~s$^{-1}$ 
velocity shift of the UV absorption lines in PG~2302+029 is potentially 
resolvable with the Advanced Imaging 
Spectrometer (ACIS), given sharp features and adequate signal-to-noise 
ratio. 

\section{Observations and Data Reduction}

We observed PG 2302+029 with \textit{Chandra} using ACIS on 7 January 2000. 
We used the most recent 
(2 November 2000) re-processed data released by the Chandra X-ray 
Center for our analysis. No filtering for high background or bad aspect 
times was required because the light curves did not show any flare-ups 
in the count rate and the aspect solution, the pattern by which 
the telescope was dithered to distribute the incoming photons over different 
pixels to minimize pixel-to-pixel variation, did not 
have any outlying points. We performed data extraction and calibration 
using version 1.4 of the Chandra Interactive Analysis of Observations (CIAO). 
We created the response matrix and ancillary response files using  
calibration data provided when the chip temperature during observations 
was $-120^{\circ}$ C.

We extracted the source counts from a circular region of radius of 
$5\arcsec$ while the background region was an annulus with radii 
between $10\arcsec$ and $20\arcsec$, both centered on the position of 
PG~2302+029. The position of PG~2302+029 
($\alpha({\rm J2000})= 23^h04^m44^s$, 
$\delta({\rm J2000})=+03\arcdeg 11\arcmin 46\arcsec$), 
as determined from the ACIS image, coincides to within $\sim 1\arcsec$ 
with the optical position of the source reported in Schneider et al. (1992). 
We obtained a total of $391\pm 21$ counts 
in an exposure time of 48 ksec across the observed energy range 
$\sim 0.4-4.0$~keV.

\section{Analysis and Results}

We bin the spectrum to have at least 30~counts/bin and use XSPEC 
(Arnaud 1996) to perform the analysis. The low count rate indicates that 
X-ray absorption may be present. We, therefore, consider fits to the data 
that include an X-ray continuum attentuated through an ionized X-ray 
absorber. Neutral X-ray absorption can be ruled out because the UV 
spectra do not contain any low-ionization metal lines that can be 
identified with the NAL and/or mini-BAL systems discussed in this 
paper (Jannuzi et al. 
1996; Jannuzi et al. 2003). We model the ionized absorbers using the 
photoionization code CLOUDY (Ferland et al. 1998) assuming solar abundances. 

We use this code to generate grids of absorbed continua in the following way: 
The incident continuum is a piecewise powerlaw, $f_\nu \propto \nu^{-\alpha}$, 
(Zheng et al. 1997; Laor et al. 1997; Telfer et al. 2002) and is displayed 
in Figure 1. The far-UV part of the spectrum is characterized by 
the 2-point spectral index $\alpha_{\rm ox}$, which relates the flux 
densities at 2500~\AA\ and 2~keV: 
$\alpha_{\rm ox}=0.384 \log(f_\nu (2500 \AA)/f_\nu (2 \rm keV))$. We use the  
B magnitude of 16.3 for PG~2302+029 to anchor the spectral energy distribution 
(SED) at 2500\AA, taking into account the appropriate $k-$correction (Green 
1996). Consequently, the X-ray flux density is specified through 
$\alpha_{\rm ox}$. We step $\alpha_{\rm ox}$ from 1.6 to 2.2 in increments 
of 0.2. This effectively gives us 4 different unabsorbed spectra representing 
4 possible intrinsic SEDs from the quasar's central engine, leading 
to 4 different intrinsic X-ray luminosities (cf. Figure 1).  
Each SED is then attenuated through an ionized absorber. The amount of 
absorption largely depends on the intrinsic total hydrogen column density, 
$N_{\rm H}$ (cm$^{-2}$), and the ionization 
parameter, $U$, defined as the ratio of the density of hydrogen ionizing 
photons to that of hydrogen particles (H$^0$ + H$^+$). For each of our 4 SEDs, 
we create a grid of attenuated continua by calculating ionized absorbers 
on a grid of $U$ and $N_{\rm H}$. Therefore, for every SED, there will be a 
grid of absorbed X-ray spectra identified by the ($U$, $N_{\rm H}$) 
combinations of the absorber through which they were transmitted.  

XSPEC incorporates a $\chi^2$ minimization scheme that allows us to find 
the best fitting attenuated continuum to the X-ray spectrum of PG~2302+029. 
We note that all our X-ray fits include attenuation by a Galactic column 
density of $N^{Gal.}_{\rm H}=5\times 10^{20}$ cm$^{-2}$ (Lockman \& Savage 
1995) and the presence of an emission line with a Gaussian profile 
at $\sim 3$ keV (probably Fe K$\alpha$ at the redshift of the QSO).  
The significance of the Fe K$\alpha$ 
detection is clearly very low, given just 391 
counts (see Figure 2), but we include it in our fits because the line is 
plausibly present and it improves the $\chi^2$ results.  
We adjust the properties of the best Fe K$\alpha$ fit--- normalization 
($1.7\times 10^{-6}$ photon s$^{-1}$ cm$^{-2}$ keV$^{-1}$), width (0.3 keV), 
and energy (2.9 keV)--- by hand after finding the best fit for the continuum 
in other parts of the spectrum. Our scheme also allows us to place the 
intrinsic absorber at any redshift we want. Naturally, we choose to place 
it either at the systemic redshift of the QSO, 
$z_{XAbs} = z_{QSO} \approx 1$, or  at the redshift of the UV absorption 
lines, $z_{Xabs}=z_{UVabs} \approx 0.7$. 

For a specific absorber redshift and for each of our 4 grids corresponding 
to the 4 intrinsic SEDs, we fit the data, with $U$ and $N_{\rm H}$ as 
free parameters, and note the value of the lowest 
reduced $\chi^2$, $\chi^2_\nu$, possible for that particular SED, and 
hence $\alpha_{\rm ox}$, and $z$. Upon comparison between the 8 
$\chi^2_\nu$ values, we find that the lowest $\chi^2_\nu \approx 1$, 
was that for 
the SED with intrinsic $\alpha_{\rm ox}=2.0$, corresponding to 
a far-UV/X-ray slope of $\alpha_{\rm EUV} = 2.4$, for both $z=0.7$ and $1.0$. 
The rest of the SEDs are rejected at the 95\% confidence level, 
regardless of the amount of intrinsic absorption. 

The need for intrinsic absorption is illustrated in Figure 2, where we
show the X-ray spectrum of PG~2302+029 together with best 
possible fits, for the continuum with $\alpha_{\rm ox}=2.0$, with no 
intrinsic absorption (upper panel), and with an ionized absorber at 
$z=0.7$ (lower panel). Clearly, the fit that includes
intrinsic ionized absorption is much better than the one that does not. 
In particular, the unabsorbed spectrum fits the data well at high 
energies, but it substantially over predicts the counts in soft X-rays. 
An X-ray absorber naturally lowers the soft X-ray flux to the measured 
values. 

Figure 3 shows overplots of the 67\%, 90\%, and 99\% confidence contours 
from fitting the data for the case where the X-ray absorber is at the 
emission redshift, $z_{XAbs} = z_{QSO}$ (solid contours),  and where 
the X-ray absorber matches the UV absorber redshift, 
$z_{XAbs} = z_{UVabs}$ (dotted contours). 
For the $z_{XAbs} = z_{QSO}$ case we find that   
$N_{\rm H} = 10^{22.8 \pm 0.2}$ cm$^{-2}$ and 
$\log U= 2.2_{-0.2}^{+0.1}$ while for the $z_{XAbs} = z_{UVabs}$,  
$N_{\rm H} = 10^{22.4 \pm 0.1}$ cm$^{-2}$ and 
$\log U= 1.6 \pm 0.3 $, where all errors are at the 90\% confidence 
level. The $\chi^2_\nu$ in both cases is $\sim 1$, and therefore the fits 
do not constrain the redshift of thte X-ray absorber. 
Note that these results were derived assuming a fixed X-ray 
powerlaw index of $\alpha_x = 0.9$. We did not explore other X-ray 
spectral slopes because the data quality is not sufficient to constrain 
additional free parameters. 

\section{Discussion}
\subsection{Intrinsic X-ray Brightness} 

Our finding in Section 3 that the intrinsic $\alpha_{\rm ox}$ is 2.0 
for PG~2302+029 deserves more attention. It has been known 
that there is a correlation between the optical luminosity at 2500\AA\ of 
a quasar and its intrinsic 
$\alpha_{\rm ox}$ (e.g., Yuan et al. 1998 and references 
therein). The relation shows a wide scatter in the distribution. 
The average $\alpha_{\rm ox}$ for radio quiet QSOs, 
such as PG 2302+029 (Kellerman et al. 1989), is 1.7 (Yuan et al. 1998; 
Vignali, Brandt, \& Schneider 2003). 
PG~2302+029, with $\alpha_{\rm ox}=2.0$ and 
$L_\nu (2500\AA) \approx 10^{31.8}$ erg s$^{-1}$ Hz$^{-1}$, is still 
consistent with the scatter in the $L_\nu-\alpha_{\rm ox}$ distribution 
discussed in Yuan et al. (1998), although it is 
$\sim 10^{\frac{2.0-1.7}{0.384}} \sim 6$ times 
fainter at $2$~keV, in the rest frame, than a quasar with comparable 
UV luminosity. This faintness is probably due to X-ray absorption.

\subsection{The UV/X-ray Relationship}

We perform a joint analysis of the our X-ray spectra and previous HST UV 
data (Jannuzi et al. 1996; Jannuzi et al. 1998; Jannuzi et al. 2003) 
to explore the relationship between the UV and X-ray absorbers. The  
HST data taken over two epochs (UT dates 19 May 1994 and 21, 24 December 
1998 ) reveal  variability in the strength of the UV absorption lines 
(Jannuzi 2002). These results confirm the intrinsic nature of the UV 
absorption, but it is important to keep in mind that the X-ray observations 
are not simulateneous with the HST dataset.

The 1994 Faint Object Spectrograph (FOS) spectra show marginally
resolved narrow absorption lines (NALs, $FWHM \approx 330$ km s$^{-1}$, 
compared to a resolution of $\approx 270$ km s$^{-1}$) of
\ion{O}{6} $\lambda\lambda 1032, 1038$, Ly$\alpha$, \ion{N}{5}
$\lambda\lambda 1239, 1243$, and \ion{C}{4} $\lambda\lambda 1548,
1551$ at $z_{abs}^{narrow}=0.7016$; broad absorption features
(mini-BALs, $FWHM \approx 3300$ km s$^{-1}$) from \ion{O}{6},
\ion{N}{5}, \ion{C}{4} are seen at $z_{abs}^{broad}=0.695$ (Jannuzi et
al. 1996). Space Telescope Imaging Spectrograph (STIS) data from 1998
reveal that the absorption, mini-BALs and NALs, in \ion{O}{6} and
\ion{N}{5} dramatically weakened to become unmeasurable, while the
mini-BALs of \ion{C}{4} also diminished in strength (Jannuzi 2002;
Jannuzi et al. 2003). The STIS spectra were obtained using the G140L and
G230LB gratings, providing coverage from 1200 to 3000 \AA\ at a spectral
resolution of roughly R$=$500, or approximately a factor of two lower than
the earlier FOS spectra. For further comparison of the STIS and FOS spectra, 
Jannuzi et al. (2003) convolved the FOS spectra to match the lower resolution
of the STIS data. The observed absorption systems were then fit with
Gaussians to determine the equivalent widths and line widths for each feature.
These measurements were used, as detailed below, to constrain the column 
densities listed in Table~1.

From the standpoint of our analysis we are dealing effectively with 4
UV absorbers and 1 X-ray absorber: the broad and narrow-line UV
absorbers of the 1994 epoch, the broad and narrow-line UV absorbers of
1998, and the X-ray absorber of 2000. We compare the properties of all
these absorbers to investigate the UV/X-ray relationship.

We present in Figure 4 contours (solid lines) of ionic column densities of 
\ion{C}{4}, \ion{N}{5}, \ion{O}{6}, and \ion{H}{1} generated by 
photoionization calculations, using the $\alpha_{\rm ox}=2.0$ SED from 
section 3, on a grid of the ionization parameter and the total hydrogen 
column density, assuming solar abundances. The closed contours in Figure 4 are 
the 90\% confidence contours calculated from the X-ray data (cf. Figure 3). 
The solid-line, closed contour is for the X-ray absorber at the redshift of 
the 
PG~2302+029, while the dotted-line one is for the X-ray absorber at the 
redshift of the UV absorber. The long-dashed and short-dashed 
contours denote the values of the ionic column densities from the broad lines 
and narrow lines, respectively, in the FOS spectra (see Jannuzi et al. 1996; 
2003 for a  detailed discussion of the UV data themselves). 
We do not show contours for the \ion{N}{5} NALs since 
the $1239$ \AA\ line appears to be weaker than the $1243$ \AA\ line, 
probably due to the low signal-to-noise ratio, contrary to atomic 
physics.

We calculated UV the 
ionic column densities through a curve-of-growth analysis (Spitzer 1978) 
using the equivalent widths and FWHMs (to estimate the Doppler $b$ 
parameters) reported in Jannuzi et al. (2003), under the assumption of 
homogenous complete coverage by the absorber of the continuum source. 
The results are listed in Table 1. The Doppler $b$ parameter is a 
convenient way of characterizing the line-of-sight velocity field; we 
do not imply that thermal  broadening is the cause of the line width. 
We include only the lines measured directly by Jannuzi et al. (2003). We do 
not list upper limits. The only mini-BALs and NALs in the STIS data that are 
discernable are those of \ion{C}{4} (see Table 1). We do not show their 
contours in Figure 4 to avoid cluttering the plots. We assume that the 
unresolved STIS NALs have the same line widths as their marginally 
resolved FOS counterparts.

We wish to make three comments here about the UV measurements. First, 
the \ion{C}{4} NAL doublets have equivalent width ratios somewhere 
between the optically thin, 2:1, and the optically thick, 1:1, limits. We 
therefore expect that these lines are not dominated by unresolved 
optically thick components, and our simple curve-of-growth analysis 
should yield reasonably accurate column densities that are, in any case, firm 
lower limits. Second, the mini-BALs are in fact 
blended doublets, implying that we have no contraints on partial 
coverage and our derived column densities for these lines are merely 
lower limits. Finally, given an equivalent width, we calculated the 
resulting ionic column density, the one shown in Figure 4. We assumed 
that the mini-BAL is due to a single transition. The oscillator strength 
of this transition is the sum of the two oscillator strengths 
of the doublet. The rest wavelength of the single transition 
is the oscillator-strength-weighted average of the rest 
wavelengths of the doublet lines. A more appropriate 
method would be to use the prescription presented in Junkkarinen, Burbidge, 
\& Smith (1983) where an effective optical depth is calculated at every point 
across the line profile taking into account the combined contributions of 
the lines in the doublet. Upon experimentation with the STIS \ion{C}{4} 
mini-BALs, we found that the 
difference between the two methods is $\sim 30$\%, within the 
measurements errors involved and making minimal difference in 
Figure 4.

While weak low-ionization UV absorption lines are seen at the redshift of the 
quasar and may be of a different nature than the NALs and mini-BALs we 
are discussing here (Jannuzi et al. 1998), high-ionization UV absorption 
lines are absent at this redshift. Moreover, there is no Lyman break, at any 
redshift, in the UV spectrum, and hence the total hydrogen column density 
in a low-ionization/neutral absorbing component is not high enough to lead 
to any observable effects on the X-ray spectrum. 
Our best fit to the X-ray absorber at $z_{Xabs} = z_{QSO}$ 
over predicts the OVI line absorption at this redshift. In particular, 
the predicted OVI column density at this redshift, $10^{15.6}$ 
cm$^{-2}$, would lead to an equivalent width of $\sim 9$ \AA\ and 
$\sim 3$ \AA\ in the observed frame if the lines have a Doppler parameter 
of $b \approx 2000$ km s$^{-1}$ and $b \approx 200$ km s$^{-1}$, 
like the high-velocity 
mini-BALs and NALs, respectively. These features should be detectable 
at $\ga$7$\sigma$ in the FOS and STIS spectra. Their absence suggests 
that the X-ray absorber is {\it not} at the quasar systemic velocity.
Note that these results do not depend on our assumption of solar metallicity, 
as long as the relative metal abundances are roughly in their solar ratios.  
The X-ray absorption is dominated by metal ions, and we use our fits 
to that to predict the strength of the metal lines in the UV. The relative
abundance of hydrogen, therefore, is not a significant factor.  

Another important constraint on the UV -- X-ray relationship 
comes from the \ion{Ne}{8} column densities (Figure 5). The 
\ion{Ne}{8} lines are at 770 and 780 \AA , within the spectral 
coverage of the STIS observations but not the FOS. The fact that they 
are high-ionization lines implies that they are good tracers of the 
X-ray gas. Our best fits to the X-ray data predict \ion{Ne}{8} 
column densities of $\sim$10$^{16.5}$ 
cm$^{-2}$ if $z_{Xabs} = z_{QSO}$ and $\sim$10$^{17.2}$ cm$^{-2}$ 
if $z_{Xabs} = z_{UVabs}$. For Doppler parameters appropriate for 
the mini-BALs or NALs, $b \approx 2000$ km s$^{-1}$ or 
$b \approx 200$ km s$^{-1}$, 
respectively, we find that the \ion{Ne}{8} column densities 
predicted by the X-rays should produce easily measurable UV lines 
($\ga40\sigma$), corresponding to observed frame equivalent widths of 
$\sim 40$ \AA\ or $\sim 3$ \AA\, respectively.  
However, there is no \ion{Ne}{8} absorption detected at either redshift.

There are several possible explanations for the apparent 
discrepancies. First, the absorber overall is complex 
and time variable. The complexity is evident from the distinct 
UV kinematic components, e.g., the NALs and mini-BALs. Also, the 
column densities derived from the UV lines do not 
define an isolated location in the $\log U$ versus $\log N_H$ plane 
(Fig.4), suggesting that the NAL and mini-BAL regions both have 
multiple zones (with different values of $U$ and $N_H$). Third, 
as we have noted above, the NALs and mini-BALs both varied between 
the 1994 and 1998 HST observations (Jannuzi 2002), and 
neither of those measurements was simultaneous with the X-ray 
data obtained in 2000. Finally, the true uncertainties in $U$ 
and $N_{\rm H}$ of the X-ray absorber are likely to be larger than 
shown in Figure 3. Those results are based on fits that fixed the 
underlying continuum shape. Letting both the continuum shape and 
absorber properties vary in the fit would clearly lead to more uncertain 
results $-$ although, without pursuing that option, it is not clear 
if the the predicted \ion{Ne}{8} absorption could be as low as the 
upper limit from the UV spectrum. 

Another complication is that the UV lines may be affected by partial 
coverage. If the absorbing gas does not fully cover the background 
light source, as we have assumed above, then there can be 
unabsorbed light filling in the 
bottoms of the UV line troughs and the column densities inferred 
from the lines will be only lower limits. Studies of other sources 
show that the coverage fraction can differ between ions and 
vary with velocity (across the line profiles) in the same ion 
(Hamann et al. 1997, Barlow \& Sargent 1997, Barlow, Sargent 
\& Hamann 1997). Our column density estimates all assume 100\% 
coverage. We can determine from the doublet ratios of the NALs that those 
lines are not optically thick absorption masked by partial coverage, 
i.e., their derived column densities should be reasonable (section 4.2). 
Moreover, we have no 
diagnostic of partial coverage for the mini-BALs. We must therefore 
keep in mind that their derived column densities are, strictly 
speaking, only lower limits. Similarily, the line strengths predicted 
from the X-ray column densities, e.g. for \ion{O}{6} and \ion{Ne}{8} 
above, are lower limits because of the assumption of 100\% coverage in 
the X-ray fits. 

In summary, the strengths of most of the UV lines are consistent with the 
X-ray measurements, if the X-ray and UV absorbers are outflowing at the 
same speed. The absence of high-ionization absorption lines at the quasar's 
redshift argues for the UV and X-ray absorbers occuring in the same general 
region at high velocity. However, the X-ray observations overpredict 
the {\it high-ionization} UV lines of \ion{Ne}{8} at all velocities. 
Simultaneous UV and X-ray observations are needed to probe 
the UV -- X-ray relationship further.

\subsection{Wind Dynamics}

If we assume the X-ray absorbing gas is outflowing with the UV absorber, 
then we can use the outflow velocity determined from the UV 
lines together with the X-ray measured total hydrogen column density to 
test the viability of radiative acceleration of the wind. Hamann (1998) 
found that the terminal velocity ($v_{terminal}$) of a radiatively 
accelerated wind is related (cf. equation 3 of Hamann 1998) 
to the total luminosity of the quasar, the 
mass of its central blackhole, the total column density of the wind 
($N_{\rm H}$), the radius at which it is launched ($R_{launch}$), and the 
fraction ($f_L$) of incident continuum energy absorbed or scattered 
by the wind along an average line of sight. 

For BALs, Hamann (1998) estimated that $f_L$ could be on the order of a few 
tenths. However, the  narrower and shallower lines in PG~2302+029 will 
intercept less continuum flux. Moreover, lower column densities compared to 
BAL flows imply that reprocessing overall is less efficient. 
We, therefore, assume that $f_L \le 0.1$ for our present analysis. We also 
adopt representative values of the blackhole mass, $10^8$ M$_\odot$, and 
luminosity, $10^{46}$~erg~s$^{-1}$, which is the Eddington luminosity 
associated with 
that mass. We assume that $v_{terminal}=56,000$~km~s$^{-1}$ and 
$N_{\rm H}=2.34\times 10^{22}$~cm$^{-2}$. These values are the 
outflow velocity of the UV 
lines (Jannuzi et al. 1996) and the column density we determined from 
the X-rays if the absorber is at $z=0.7$ (see end of section 3). 
Substituting all these numbers into equation (3) of 
Hamann (1998), we find that the launch radius is 
$\le 10^{15}$ cm $\approx 100$ R$_{Schwarzchild}$. 
Therefore, if the wind is radiatively accelerated, it must be 
launched very near to the black hole.

The mass loss rate implied by this radius is 
$\le 0.1~Q$ M$_\odot$ yr$^{-1}$, where $Q$ is the 
global covering factor $\Omega/4\pi$ (Hamann \& Brandt 2002, in preparation). 
The density of the flow at this launch radius should be 
$n_H\ga 10^{11}$ cm$^{-3}$, if it is photoionized with 
$\log U = 1.6$ appropriate for the X-ray absorber (see 
Hamann \& Brandt 2002 for explicit equations).

\subsection{Geometry and Physical Models}

The small launch radius required for a high-velocity X-ray absorber 
may be problematic for models of the outflow in PG~2302+029. 
For comparison, this maximum launch radius is {\it much} 
smaller than the nominal radius of the broad emission line region for 
a quasar of the same luminosity ($R_{\rm BLR}\approx 2\times 10^{18}$ cm, 
based on reverberation studies, e.g., Kaspi et al. 2000). On the 
other hand, the size of the X-ray continuum source is 
$\sim 10^{14}$~cm (e.g., Peterson 1993), an order of magnitude smaller  
than the launch radius derived above. Given the small launch radius, 
it seems likely that 
the X-ray absorber is either {\it not} radiatively accelerated, or {\it not} 
outflowing at the same high speed as the UV lines. 

In either case, the absence of high-ionization 
UV absorption near the systemic velocity 
places another important constraint on wind models. If it is a steady-
state flow, then the acceleration must occur in a region that 
does not intersect our sightline to the continuum source. Models of 
outflows that lead to such UV lines have been discussed by 
Murray et al. (1995), Murray \& Chiang (1995), and Elvis (2000). In 
these scenarios, the wind is initially perpendicular to the accretion 
disk. As it flows farther from the disk, 
the radiation pressure accelerating it 
bends and flares in the radial direction. Oblique lines of sight then 
could pass through the bent part of the wind, thus explaining the absence 
of zero-velocity absorption. 

The models mentioned in the previous paragraph differ in subtle ways. 
Murray et al. (1995) described a case in which the X-ray absorber is 
at rest. Its function is to shield the UV gas from soft X-rays and 
prevent it from becoming highly ionized, in which case resonant line 
driving would not be effective. Another variant of this scheme is the 
possibility of a self-shielding wind (Murray \& Chiang 1995; Elvis 2000). 
The X-ray and UV absorption arise in the same outflowing gas. 
The high column density of the wind requires a small launch radius,   
although probably larger than the radius we derive in section 4.3. 

\section{Conclusions}
We presented  
{\it Chandra} X-ray observations of the radio-quiet QSO
PG~2302$+$029 and demonstrated the presence of soft X-ray absorption in
its spectrum.  Older UV spectra of this quasar have been used to
identify the presence of rarely observed ultra-high velocity ($-56,000$
km~s$^-1$) absorption lines consistent with this quasar containing a
remarkable outflow from its active nucleus (Jannuzi et al. 1996,
Jannuzi et al. 2003).  Using photoionization models and the combined
X-ray and UV data sets we have investigated the possible physical
properties of the gas producing the X-ray and UV absorption.  We
suggest that the X-ray absorption also occurs at high velocities in the
same general region as the UV absorber.  There are not enough constraints to 
rule out multi-zone models.  Multi-zone models are required if the 
distinct broad and narrow UV line profiles are both intrinsic to the QSO.
The properties of the X-ray and UV absorption, as inferred from the 
data, are consistent with each other, if the X-ray absorbing gas is 
outflowing with the UV absorber. However, the X-ray data over predict 
the strength of the high-ionization UV lines of \ion{Ne}{8} at all 
velocities. If we assume the X-ray absorbing gas is in an outflow with 
the same velocity as the gas producing the UV-absorbing wind and that such 
winds are radiatively accelerated, then the outflow must be started at a 
radius of $\le 10^{15}$ cm from the central source of the radiation.

\noindent {\it Acknowledgements:} We wish to acknowledge support 
through {\it Chandra} grants GO 0-1123X and GO 0-1157X. BTJ acknowledges 
support from the National Science Foundation through their support of the 
National Optical Astronomy Observatory, which is operated by the Association 
of Universities for Research in Astronomy, Inc. (A.U.R.A.) under cooperative 
agreement with the National Science Foundation and from NASA through a 
grant to proposal GO-07348.01-A from the Space Telescope Science Institute, 
which is operated by A.U.R.A., Inc., under NASA contract NAS5-26555

\clearpage
\begin{figure}
\vbox{
\centerline{
\psfig{figure=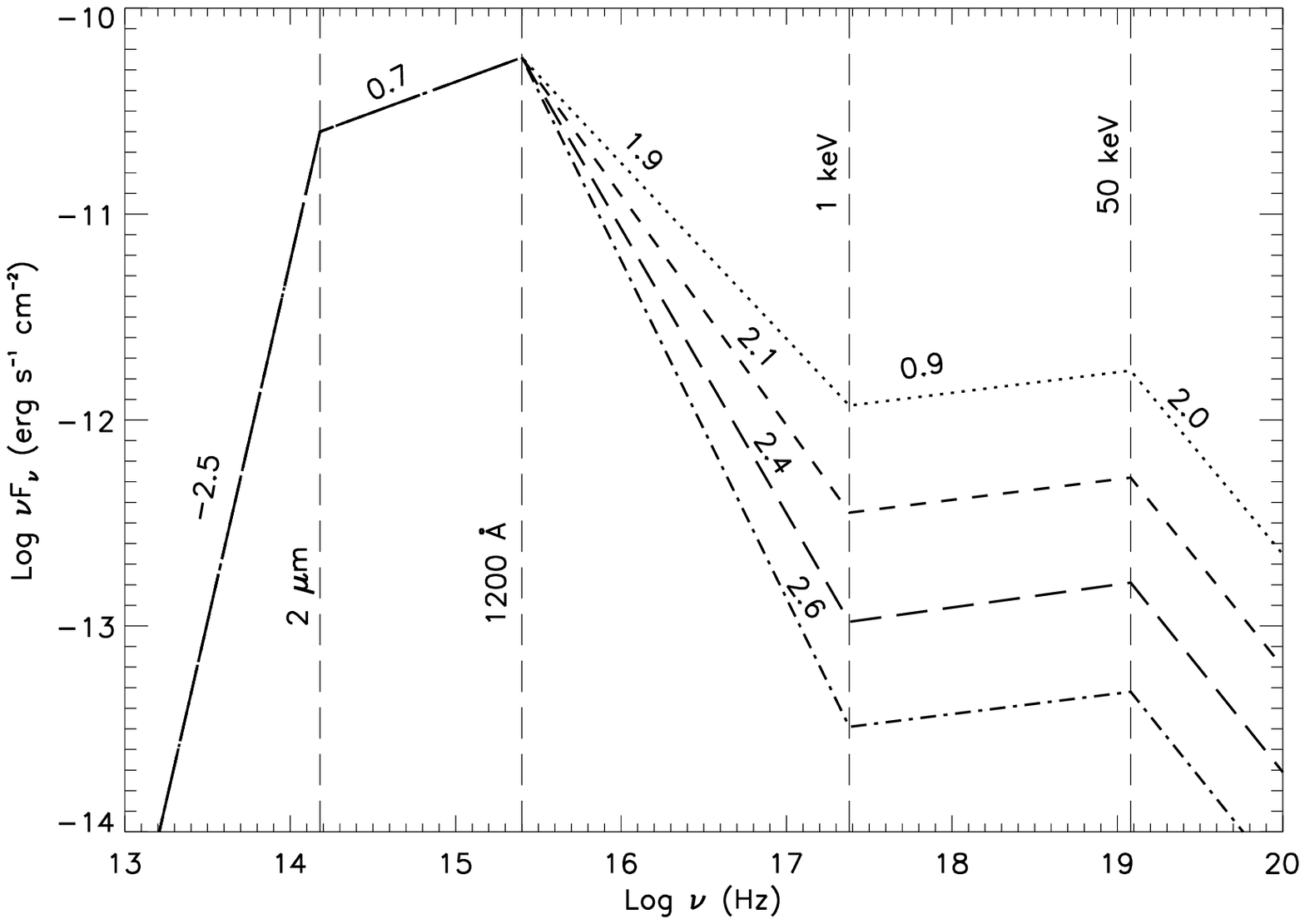}
}
\caption{Incident piece-wise powerlaw continua used in generating the grid of 
ionized absorbers. The slopes are indicated ($f_\nu \propto \nu^{-\alpha}$). 
The dotted, dashed, long-dashed, and dash dotted lines correspond to 
$\alpha_{\rm ox} = 1.6, 1.8, 2.0$, and $2.2,$ respectively.}
}
\end{figure}

\newpage
\begin{figure}
\vbox{
\centerline{
\psfig{figure=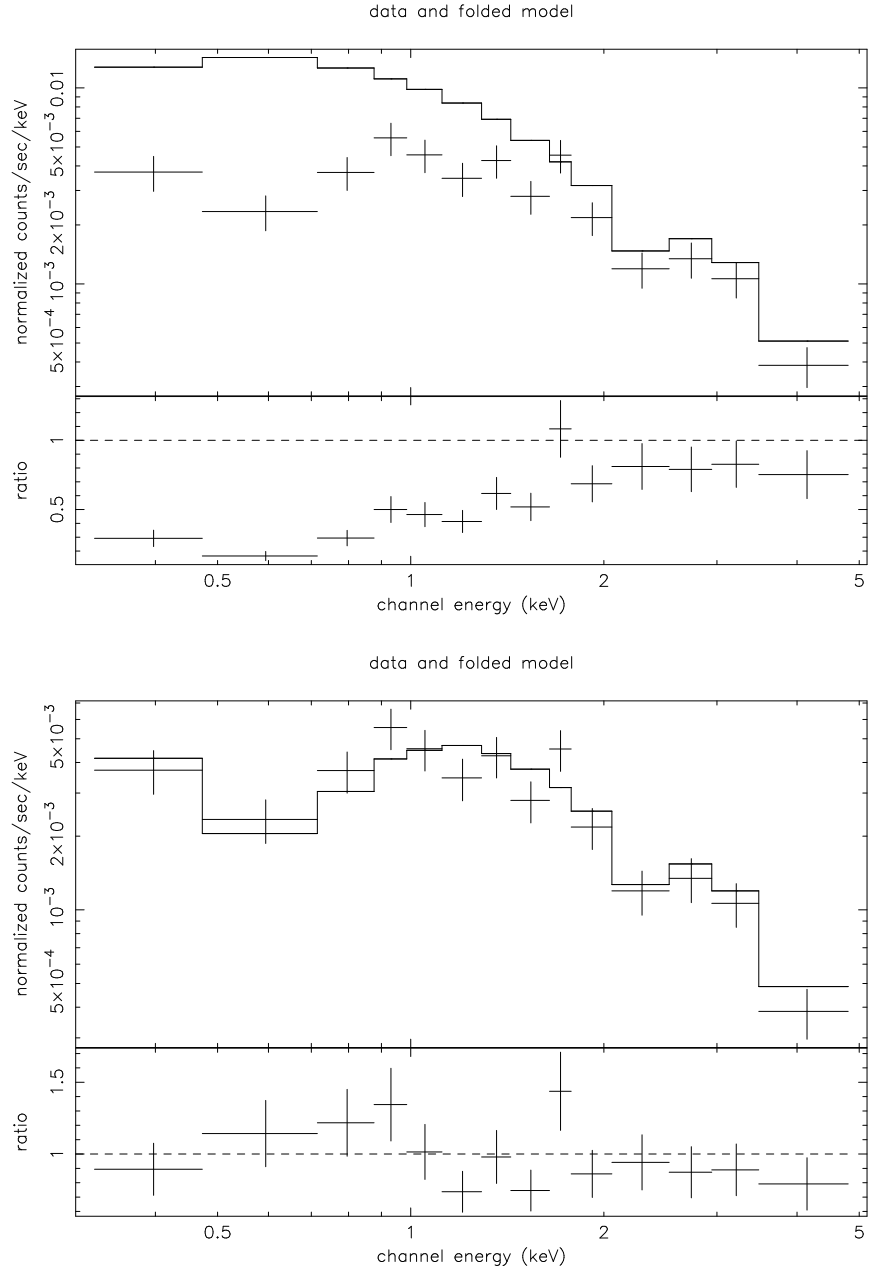}
}
\caption{X-ray spectrum of PG~2302+029. Upper Panel: with no intrinsic
absorption.  
Lower Panel: ionized absorber at $z \approx 0.7$. In both cases 
$\alpha_{\rm ox} = 2.0$}
}
\end{figure}

\newpage
\begin{figure}
\vbox{
\centerline{
\psfig{figure=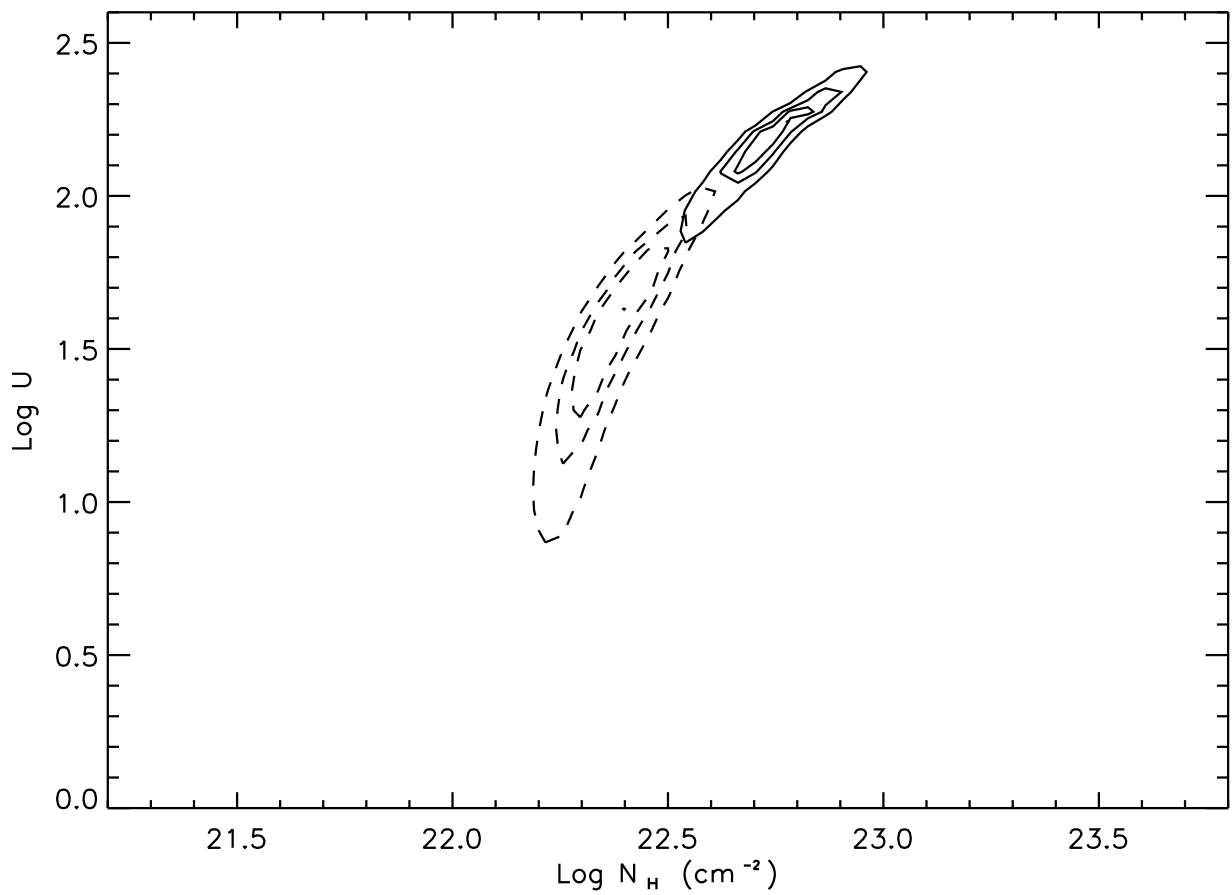}
}
\caption{Confidence contours, 67\%, 90\%, and 99\%, levels for 
X-ray fits. Solid-line contours are for $z_{Xabs}=z_{QSO}$, while 
dashed-line ones are for $z_{XAbs}=z_{UVabs}$.}
}
\end{figure}

\newpage
\begin{figure}
\vbox{
\centerline{
\psfig{figure=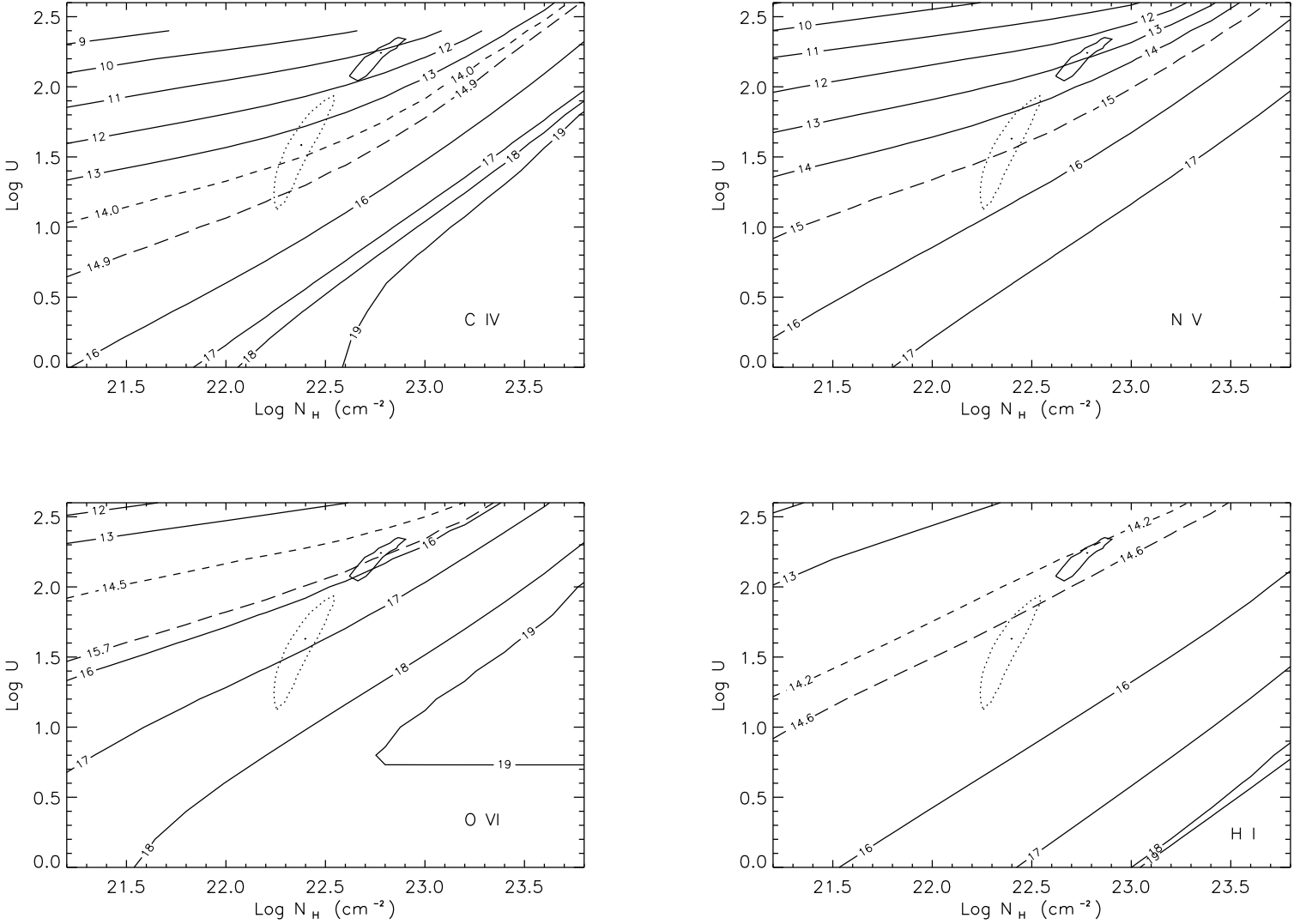}
}
\caption{Ionic column density contours as a function of ionization 
parameter and total hydrogen column density. Contours are labelled 
by the logarithm of column density expressed in cm$^{-2}$, for the 
indicated species. 
Closed contours (solid: 
$z_{Xabs}=z_{QSO}$, dotted: $z_{XAbs}=z_{UVabs}$) are the 90\% 
confidence levels determined from the fits to the X-ray data. The solid 
lines are generated from photoionization calculations. The long-dashed 
and short-dashed lines are the column densities measured from the 
FOS spectra for the broad and narrow absorption lines, 
respectively. The long-dashed contour for \ion{H}{1} is an upper limit 
on broad Ly$\alpha$ absorption assuming complete line-of-sight 
coverage (Hamann 1997). 
We do not show the measurement of the narrow 
N V lines, since the line ratio does not make sense physically.}
}
\end{figure}

\newpage
\begin{figure}
\vbox{
\centerline{
\psfig{figure=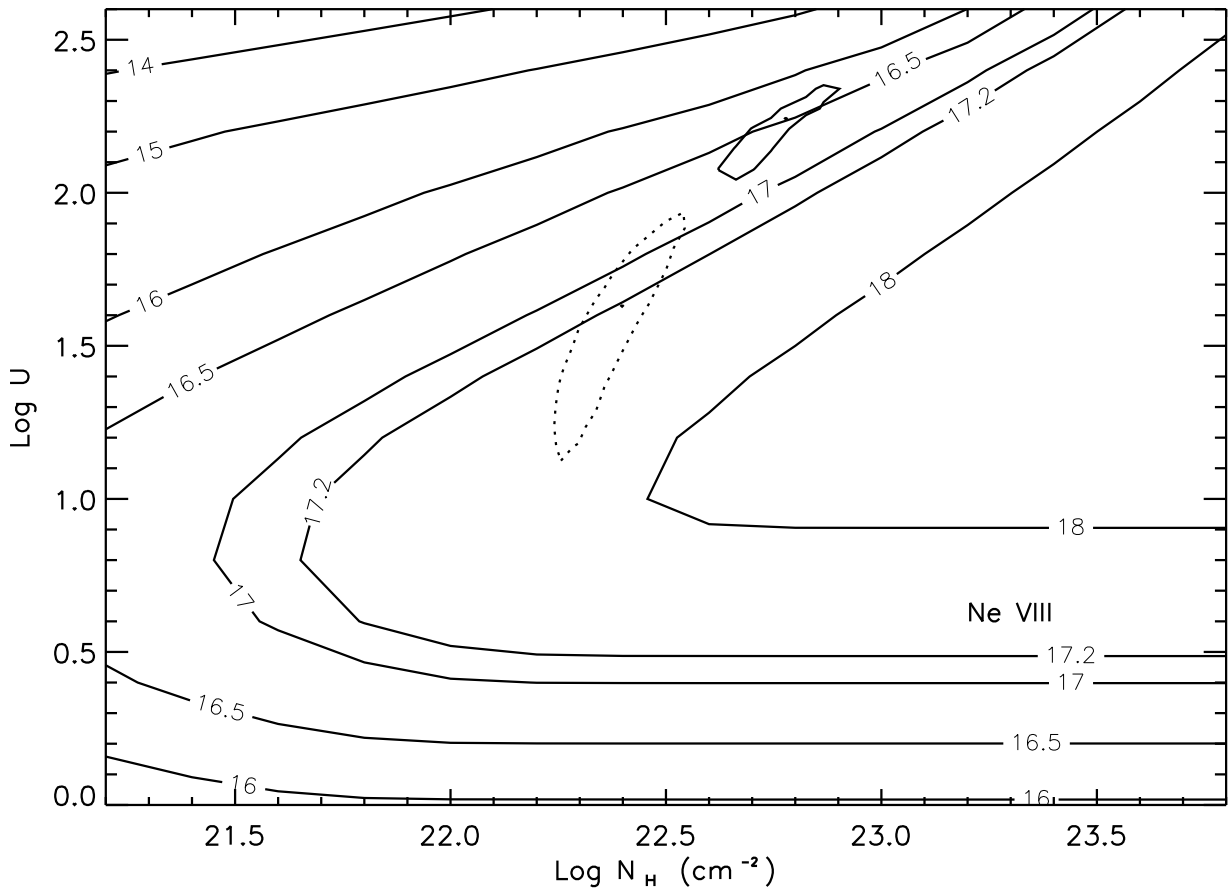}
}
\caption{Column density contours of \ion{Ne}{8}, as calculated with 
CLOUDY. Contour conventions are the same as in Figure 4.}
}
\end{figure}

\clearpage

\begin{table}
\begin{center}
\title{\rm \small Table 1: Adopted UV Absorber Parameters \rm}
\begin{tabular}{ccccc}
\\
\tableline
\tableline
From FOS Observations of 1994\\
Ion &Wavelength (\AA)$^1$ &b (km~s$^{-1}$) & W (\AA)$^1$ & N$_{ion}$ (cm$^{-2}$)\\
\tableline
\ion{C}{4} & $1548+1551$ & 2060 & 4.24 & $10^{14.9}$\\
\ion{O}{6} & $1032+1038$ & 2146 & 7.08 & $10^{15.7}$\\
\ion{N}{5} & $1239+1243$ & 1854 & 3.10 & $10^{15.0}$\\
\ion{C}{4} & 1548       & 200  & 0.31  & $10^{14.0}$\\ 
\ion{O}{6} & 1031       & 330  & 0.37 & $10^{14.5}$\\
\ion{H}{1} & 1216       & 150  & 0.63 & $10^{14.2}$\\
\tableline
From STIS Observations of 1998\\
Ion &Wavelength (\AA)$^1$ &b (km~s$^{-1}$) &  W (\AA)$^1$ & N$_{ion}$ (cm$^{-2}$)\\
\tableline
\ion{C}{4} & $1548+1551$ & 2080     & 3.40 & $10^{14.8}$\\
\ion{C}{4} & 1548      & 200$^2$  & 0.30& $10^{14.0}$\\
\tableline
\end{tabular}
\end{center}
$^1$ Rest frame \\
$^2$ In the STIS spectrum this line is unresolved. We assumed its profile and then 
assumed that the $b$ parameter matched the value measured in the 
earlier FOS spectrum. 
\end{table}

\end{document}